\documentclass[conference]{IEEEtran}
\IEEEoverridecommandlockouts

\usepackage{cite}

\ifCLASSINFOpdf
\else
\fi
\usepackage[cmex10]{amsmath}
\usepackage{algorithmic}
\usepackage{array}
\usepackage{mdwmath}
\usepackage{eqparbox}
\usepackage{amssymb}
\usepackage{amsfonts}
\usepackage{graphicx,amssymb,lineno}
\usepackage{algorithm}
\usepackage{algorithmic}
\usepackage[usenames]{color}

\begin{document}

\title{Frequency-Domain Response Based Timing Synchronization: A Near Optimal Sampling Phase Criterion for TDS-OFDM}%

%\author{
%\authorblockN{Zhen Gao, Chao Zhang, Yu Zhang}
%\authorblockA{Tsinghua National Laboratory for Information \\ Science and Technology (TNList),\\
% Department of Electronic Engineering, \\ Tsinghua University, Beijing 100084, P. R. China,\\
% National Engineering Lab. for DTV (Beijing)\\
% Email: gao-z11@mails.tsinghua.edu.cn}
%\and
%\authorblockN{Hang Zhang}
%\authorblockA{Science and Technology on Information Transmission \\ and Dissemination in Communication Networks Laboratory}
%}

\author{\IEEEauthorblockN{Zhen Gao\IEEEauthorrefmark{1}, Chao Zhang\IEEEauthorrefmark{1}, Yu Zhang\IEEEauthorrefmark{1} and Hang Zhang\IEEEauthorrefmark{2}}
\IEEEauthorblockA{\IEEEauthorrefmark{1}Tsinghua National Laboratory for Information Science and Technology (TNList),\\
 Department of Electronic Engineering, Tsinghua University, Beijing 100084, P. R. China,\\
 National Engineering Lab. for DTV (Beijing)\\
 Email: gaozhen010375@foxmail.com}
\IEEEauthorblockA{\IEEEauthorrefmark{2}Science and Technology on Information Transmission and
 Dissemination in Communication Networks Laboratory}
}

\maketitle

\begin{abstract}
%\boldmat
In time-domain synchronous OFDM (TDS-OFDM) system for digital television terrestrial multimedia broadcasting (DTMB) standard, the baseband OFDM signal is upsampled and shaping filtered by square root raised cosine (SRRC) filter before digital-to-analog converter (DAC). Much of the work in the area of timing synchronization for TDS-OFDM focuses on frame synchronization and sampling clock frequency offset recovery, which does not consider the sampling clock phase offset due to the upsampling and SRRC filter. This paper evaluates the bit-error-rate (BER) effect of sampling clock phase offset in TDS-OFDM system. First, we provide the BER for $M$-order quadrature amplitude modulation ($M$-QAM) in uncoded TDS-OFDM system. Second, under the condition of the optimal BER criterion and additive white Gaussian noise (AWGN) channel, we propose a near optimal sampling phase estimation criterion based on frequency-domain response. %Compared with conventional time-domain based timing synchronization method in TDS-OFDM, the proposed criterion
Simulations demonstrate that the proposed criterion also has good performance in actual TDS-OFDM system with channel coding over multipath channels, and it is superior to the conventional symbol timing recovery methods for TDS-OFDM system.
\end{abstract}



% \fi
\IEEEpeerreviewmaketitle

\section{Introduction}
\label{sec:intro}
%先说1ofdm 使用over src用处，同时dtmb就是这样，
%2 针对offset也有一些其他工作，比如。。但是我们不同于他们，
%3 DTMB有一种
%4 我们的
%5
Time-domain synchronous OFDM (TDS-OFDM) has superior performance in terms of fast synchronization, accurate channel estimation and higher spectral efficiency compared with other OFDM solutions, and it has been adopted by the digital television terrestrial multimedia broadcasting (DTMB) standard \cite{{general},{sparse2}}. In TDS-OFDM system, the baseband OFDM signal is upsampled and shaping filtered by the square root raised cosine (SRRC) filter before digital-to-analog converter (DAC) \cite{GB2006}. In this way, the spectrum outside the band is effectively suppressed, the inter-symbol-interference within an OFDM data block is degraded, and the correlation based frame synchronization can be robust to carrier frequency offset (CFO) and multipath channels \cite{{FS0},{FS1},{FS2},{FS3}}.

Previous work in the area of timing synchronization for TDS-OFDM focuses on frame synchronization and sampling clock frequency offset correction. \cite{FS0} proposed a symbol timing recovery (STR) method based on code acquisition (CA) to obtain frame synchronization and track the sampling clock frequency offset at the receiver. %This method exploits the time-domain pseudo noise (PN) sequence correlation to and focuses on the the extraction of timing error which drives the phase-locked loop (PLL) to track the sampling frequency offset
However, this CA based frame synchronization suffers from obvious performance loss when large CFO exists. Hence \cite{{FS1},{FS2},{FS3},{FS4}} proposed robust frame synchronization methods for TDS-OFDM. However, the STR method proposed in \cite{{FS0},{FS1},{FS2},{FS3},{FS4}} cannot obtain the optimal sampling clock phase over multipath channels, which will be discussed in this paper.
In respect of timing synchronization in other communication systems, \cite{chedan1} investigated the effect of frame synchronization error in general OFDM systems. \cite{COFDM} proposed a pilot-aided sampling frequency offset recovery method in cyclic prefix OFDM (CP-OFDM).
\cite{chedan2} and \cite{chedan3} examined the effect of clock jitter in cooperative space-time coding multiple input single output systems (MISO).

In contrast, to the best of our knowledge, this is the first paper to investigate the impact of sampling clock phase offset on system performance for TDS-OFDM system, which is superior to other OFDM solutions and has been adopted by DTMB standard. In this paper, we evaluate the effect of sampling clock phase offset owing to upsampling and SRRC filter shaping in the TDS-OFDM system after the perfect frame synchronization, sampling clock frequency offset recovery, and CFO elimination. Meanwhile, we also propose a near optimal sampling phase estimation criterion based on frequency-domain response, which is different from the conventional time-domain based timing synchronization methods for TDS-OFDM.
% Our work differs from the previous work in that we investigate the effect of sampling phase offset in the TDS-OFDM based on perfect sampling frequency offset and frame synchronization.

This paper focuses on three problems as follows. Whether the sampling phase offset does have a great influence on BER performance or not. If yes, is there an optimal or near optimal criterion to solve this problem? Compared with the conventional synchronization methods, %how about the performance superiority of the proposed criterion?
how much performance gain can be achieved by the proposed criterion?

%In this paper, we first indicate that the sampling phase offset does have great influence on the bit-error-rate (BER) performance in TDS-OFDM system. Then, we provide the BER with different sampling phase offsets in uncoding TDS-OFDM, and also propose a near optimal sampling phase criterion under the condition of additive white gaussian noise (AWGN) channel. Simulations demonstrate that the proposed sampling phase criterion performs well in actual TDS-OFDM with channel coding, and it is superior to conventional time recovery method \cite{FS0}.

The rest of the paper is organized as follows. In Section II, we introduce the baseband model of DTMB system and the conventional synchronization methods for TDS-OFDM. Meanwhile, the effect of sampling phase offset in TDS-OFDM is presented. In Section III, we provide the BER of uncoded TDS-OFDM system and propose a near optimal sampling phase estimation criterion. In Section IV, simulation results are provided. In Section V, conclusions are drawn.

\section{System Model of TDS-OFDM Based DTMB }
\begin{figure}[tblp]
     \centering
     \includegraphics[width=3.4cm, keepaspectratio,angle =90]
     {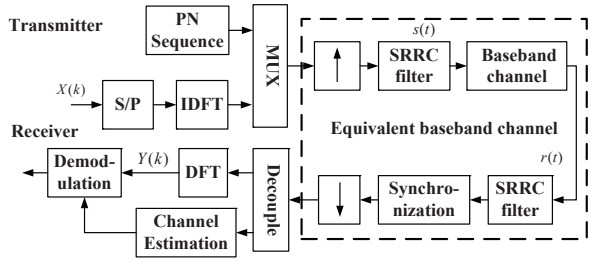}
     \caption{Baseband transceiver of the TDS-OFDM system.}
     \label{fig:baseband}
    \vspace*{-4mm}
\end{figure}

The baseband transceiver of the TDS-OFDM system \cite{GB2006} is shown in Fig. \ref{fig:baseband}. In the time domain, a TDS-OFDM symbol consists of a pseudo-noise (PN) sequence and the following OFDM data block. The PN sequence, serving as the guard interval, is inserted between the adjacent OFDM data blocks to eliminate the inter-block-interference (IBI) over multipath channels. The OFDM data block is generated by inverse discrete Fourier transform (IDFT) of frequency-domain data. Both the PN sequence and the OFDM data block share the same symbol rate $f_{\rm{sym}}=1/T_{\rm{sym}}=7.56\text{MHz}$. After multiplexing, the TDS-OFDM signal is processed by $N_{\text{upsam}}=4$ times upsampling and SRRC filter shaping, thus the signal sampling rate becomes $f_{\text{upsam}}=30.24\text{MHz}$. Finally, the signal is sent to DAC.

At the receiver, the baseband sampling rate of the analog-to-digital converter (ADC) is $f_s=1/T_s=30.40\text{MHz}$, which is slightly higher than $f_{\text{upsam}}$ to reduce the baseband signal information loss in the absence of synchronization \cite{{WJPhd},{DCR}}. The synchronization module is aimed at frame synchronization, CFO elimination and the sampling clock frequency recovery. Sequentially, signal after synchronization is downsampled, and then PN and OFDM data block are decoupled, whereby PN is used for channel estimation and OFDM data block is sent to equalization and demodulation.
\begin{figure}[blp]
     \centering
     \includegraphics[width=3.05cm, keepaspectratio,angle =90]
     {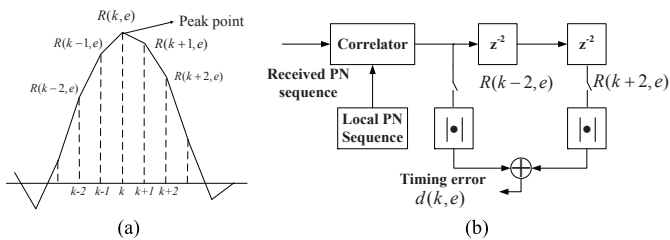}
     \caption{The conventional STR method in TDS-OFDM: (a) time-domain correlation; (b) timing error detector.}
     \label{fig:STR}
     \vspace*{-5mm}
\end{figure}

The synchronization module consists of three parts: frame synchronization, CFO correction and sampling frequency offset recovery. \cite{FS0} proposed a STR method based on CA for TDS-OFDM. In CA stage, this method searches and tracks the correlation peak $R(k,e)$ of the PN sequences embedded in the signals to obtain the frame synchronization, where $k$ is the time index of the correlation peak and $e$ is the timing error, as shown in Fig. \ref{fig:STR}. (a). In this way, the timing error is reduced to no more than $\pm {T_{\text{sym}}}/2$. Sequentially, the STR algorithm eliminates the residual timing error $e$ by a STR feedback loop, which consists of a timing error detector, a loop filter, and a digital interpolator. The interpolator driven by the timing error signal is used to recover the received signal, and the loop filter is used to normalize the timing error signal and enhance its robustness to noise. Timing error signal is produced by the amplitude difference of adjacent sidelobes of the acquired maximum correlation peak, as shown in Fig. \ref{fig:STR} (b). Consequently, signal after the interpolator is adjusted to sampling frequency $f_{\text{upsam}}=30.24\text{MHz}$ by a decimator.

Nevertheless, the STR method based on the time-domain PN correlation aims at tracking the sampling clock frequency offset, and it cannot obtain the optimal sampling clock phase offset over multipath channels. %The following down-sampling extracts the samples of the signal after synchronization module every other $N_{\text{upsam}}-1$ samples, which is based on the the acquired maximum correlation peak.
Therefore, some questions appear. Does the sampling clock phase offset have a great influence on the BER performance? If yes, is there an optimal or near optimal criterion to solve this problem? Compared with the conventional STR methods, how much performance gain can be achieved by the proposed criterion?

\begin{figure}[tblp]
     \centering
     \includegraphics[width=8.6cm, keepaspectratio]
     {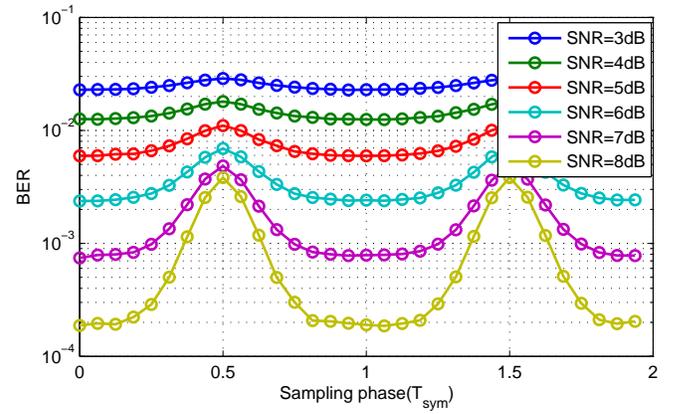}
     \caption{BER against different sampling phase offsets for TDS-OFDM using uncoded BPSK modulation over AWGN channel.}
     \label{fig:bpsk_phase}
     \vspace*{-4mm}
\end{figure}

For the first problem, we provide the BER of TDS-OFDM system over additive white Gaussian noise (AWGN) channel with different sampling phase offsets, as shown in Fig. \ref{fig:bpsk_phase}. Simulation assumes perfect frame synchronization, sampling clock frequency offset recovery, and CFO elimination. OFDM data block adopts uncoded binary phase shift keying (BPSK) modulation with DFT length $N=4096$, $N_{\text{upsam}}=4$, and SRRC roll-off factor $\alpha=0.05$. From Fig. \ref{fig:bpsk_phase}, we observe that the BER performance of different sampling phase offsets within a $T_{\text{sym}}$ varies largely, and it appears $T_{\text{sym}}$ periodicity. Even the optimal sampling phase offset is superior to the worst situation by about 3 dB performance gain.

\section{The Proposed Sampling Phase Offset Criterion}
\label{proposed}
In this section, we provide the BER in uncoded TDS-OFDM system, and propose a near optimal sampling phase estimation criterion.
\subsection{Frequency-Domain Response of The Equivalent Baseband Channel}
In Fig. \ref{fig:baseband}, we consider modules in the dashed box as the equivalent baseband channel. We assume that signal after synchronization module achieves perfect frame synchronization, sampling clock frequency offset recovery, and CFO elimination. The analog frequency-domain response of SRRC filters, including the transmit and receiver shaping filters, is denoted as ${H_{{\rm{SRRC}}}}(\Omega )$, which can be expressed as
\begin{align}
{H_{{\rm{SRRC}}}}(\Omega ) = \left\{ \begin{array}{l}
 {T_{{\rm{sym}}}},{\kern 1pt} {\kern 1pt} {\kern 1pt} {\kern 1pt} {\kern 1pt} {\kern 1pt} {\kern 1pt} {\kern 1pt} {\kern 1pt} {\kern 1pt} {\kern 1pt} {\kern 1pt} {\kern 1pt} {\kern 1pt} {\kern 1pt} {\kern 1pt} {\kern 1pt} {\kern 1pt} {\kern 1pt} {\kern 1pt} 0 \le \left| {\left. \Omega  \right|} \right. < \frac{{(1 - \alpha )\pi }}{{{T_{{\rm{sym}}}}}}, \\
 \frac{{{T_{{\rm{sym}}}}}}{2}[1 + \sin (\frac{{{T_{{\rm{sym}}}}}}{{2\alpha }}(\frac{\pi }{{{T_{{\rm{sym}}}}}} - \Omega ))], \\
{\kern 20pt} {\kern 1pt} {\kern 1pt} {\kern 1pt} {\kern 1pt} {\kern 1pt} {\kern 1pt} {\kern 1pt} {\kern 1pt} {\kern 1pt} {\kern 1pt} {\kern 1pt} {\kern 1pt} {\kern 1pt} {\kern 1pt} {\kern 1pt} {\kern 1pt} {\kern 1pt} {\kern 1pt} {\kern 1pt} {\kern 1pt} {\kern 1pt} {\kern 1pt} {\kern 1pt} {\kern 1pt} \frac{{(1 - \alpha )\pi }}{{{T_{{\rm{sym}}}}}} \le \left| {\left. \Omega  \right|} \right. < \frac{{(1 + \alpha )\pi }}{{{T_{{\rm{sym}}}}}}, \\
 0, {\kern 5pt} {\kern 1pt} {\kern 1pt} {\kern 1pt} {\kern 1pt} {\kern 1pt} {\kern 1pt} {\kern 1pt} {\kern 1pt} {\kern 1pt} {\kern 1pt} {\kern 1pt} {\kern 1pt} {\kern 1pt} {\kern 1pt} {\kern 1pt} {\kern 1pt} {\kern 1pt} {\kern 1pt} {\kern 1pt} {\kern 1pt} {\kern 1pt} {\kern 1pt} {\kern 1pt} {\kern 1pt} {\kern 1pt} {\kern 1pt} {\kern 1pt} {\kern 1pt} {\kern 1pt} \left| {\left. \Omega  \right|} \right. > \frac{{(1 + \alpha )\pi }}{{{T_{{\rm{sym}}}}}}. \\
 \end{array} \right.
\label{equ:HSRRC}
\end{align}
The analog frequency-domain response of baseband channel is denoted as ${H_C}(\Omega )$.%, and it can be written as ${H_C}(\Omega ) = \int\limits_{ - \infty }^\infty  {{h_C}(t){e^{ - j2\pi \Omega t}}dt}$, where $h_C(t)$ is the channel time-domain impulse response.

\begin{figure}[tblp]
     \centering
     \includegraphics[width=9cm, keepaspectratio]
     {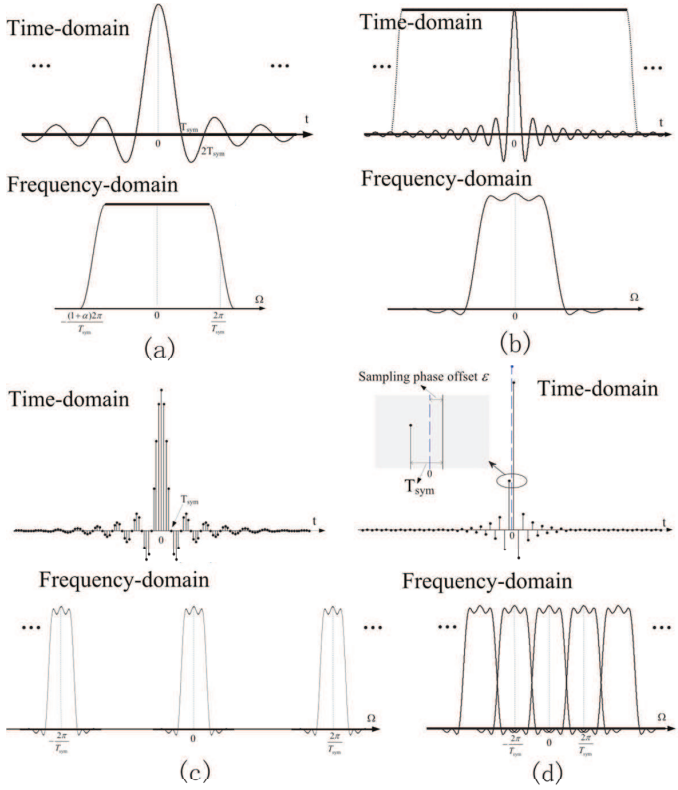}
     \caption{Frequency-domain response and the corresponding time-domain response of the equivalent baseband channel. For convenience, we consider the AWGN channel, i.e. ${H_C}(\Omega )=1$. (a) ${H_C}(\Omega ){H_{{\rm{SRRC}}}}(\Omega )$; (b) ${H_{{\rm{finite}}}}(\Omega )$; (c) ${H}'(\Omega )$; (d) ${H}(\Omega )$.}
     \label{fig:process}
     \vspace*{-5mm}
\end{figure}

%For convenience, we consider the AWGN channel, i.e. ${H_C}(\Omega )=1$ or $h_C(t)=\delta(t)$, where $\delta ( \bullet )$ is a generalized function on the real number field that is zero everywhere except at zero, with an integral of one over the entire real number field.
Fig. \ref{fig:process}. (a) illustrates the frequency-domain and its corresponding time-domain response of ${H_C}(\Omega ){H_{{\rm{SRRC}}}}(\Omega )$ under AWGN channel.
In practice, the length of SRRC filters at the transmitter and receiver is finite, which means an equivalent time-domain rectangle windowing on the response, as shown in Fig. \ref{fig:process}. (b). Since the window length is usually very long, the effect of equivalent time-domain windowing can be negligible. Consequently, the frequency-domain response after windowing is ${H_{{\rm{finite}}}}(\Omega ) \approx {H_{{\rm{SRRC}}}}(\Omega ){H_C}(\Omega )$.

Next, the impact of upsampling at the transmitter on ${H_{{\rm{finite}}}}(\Omega )$ is that the spectrum becomes periodical and compressed, as illustrated in Fig. \ref{fig:process}. (c). The spectrum after upsampling can be expressed as
\begin{align}
\begin{array}{l}
 H'(\Omega ) =  \sum\limits_{k =  - \infty }^{ + \infty } {\frac{{{H_{{\rm{finite}}}}(\Omega  - \frac{{2\pi k}}{{{T_{{\rm{sym}}}}}})}}{{{N_{{\rm{upsam}}}}{T_{{\rm{sym}}}}}}}  \approx  \\
{\kern 10pt}  \sum\limits_{k =  - \infty }^{ + \infty } {\frac{{{H_C}({N_{{\rm{upsam}}}}(\Omega  - \frac{{2\pi k}}{{{T_{{\rm{sym}}}}}})){H_{{\rm{SRRC}}}}({N_{{\rm{upsam}}}}(\Omega  - \frac{{2\pi k}}{{{T_{{\rm{sym}}}}}})}}{{{N_{{\rm{upsam}}}}{T_{{\rm{sym}}}}}}}. \\
 \end{array}\label{equ:H10}
\end{align}

Finally, the influence of downsampling at the receiver on $H'(\Omega )$ is spectrum aliasing, as illustrated in Fig. \ref{fig:process}. (d). And the frequency-domain response of final equivalent baseband channel can be written as
\begin{align}
\!\!\!\!\!\!\!\!\!\!\!\!\!\!\!\!\!\!&H(\Omega ) \approx \nonumber \\
&\sum\limits_{k =  - \infty }^{ + \infty } {\frac{{{H_C}(\Omega  - \frac{{2\pi k}}{{{T_{{\rm{sym}}}}}}){H_{{\rm{SRRC}}}}(\Omega  - \frac{{2\pi k}}{{{T_{{\rm{sym}}}}}}){e^{j(\Omega  - \frac{{2\pi k}}{{{T_{{\rm{sym}}}}}})\varepsilon {T_{{\rm{sym}}}}}}}}{{{T_{{\rm{sym}}}}}}} ,\label{equ:H1}
\end{align}
where $\varepsilon  \in [ - 0.5,0.5]$ is normalized sampling phase offset.
\begin{figure}[tblp]
     \centering
     \includegraphics[width=8cm, keepaspectratio]
     {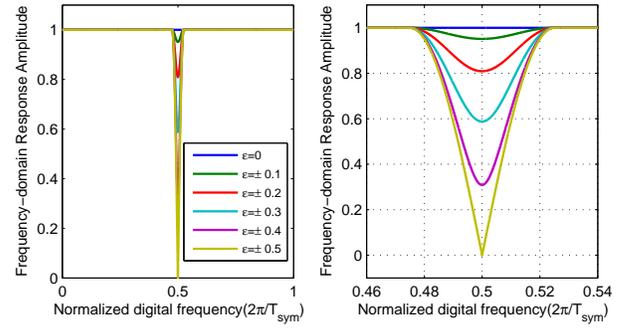}
     \caption{Frequency-domain response amplitude of $H(\Omega )$ with different sampling phase offsets. Here, we consider the AWGN channel, i.e. ${H_C}(\Omega )=1$, and $\alpha=0.05$. The right figure is the details of the left figure in $f \in [0.46,0.54]$.}
     \label{fig:freq_vs_e}
     \vspace*{-5mm}
\end{figure}

Since $H(\Omega )$ appears $T_{\text{sym}}$ periodical, we only need investigate $H(\Omega )$ of $\Omega  \in [0,2\pi /{T_{{\rm{sym}}}}]$ or $f  \in [0,1]$, where $f  = \frac{{\Omega {T_{{\rm{sym}}}}}}{{2\pi }}$ is the normalized digital frequency. Fig. \ref{fig:freq_vs_e} provides the frequency-domain response amplitude of $H(\Omega )$ with different sampling phase offsets, where ${H_C}(\Omega )=1$. It can be observed that with the sampling phase offset increasing, $|H(\Omega )|$ near $f=0.5$ rapidly decreases. Obviously, sampling phase offset $\varepsilon$ has a great influence on $|H(\Omega )|$.
\subsection{BER and The Proposed Sampling Phase Criterion.}
According to \cite{BER_MQAM}, the uncoded symbol-error-rate (SER) of $M$-order quadrature amplitude modulation ($M$-QAM) over AWGN channel is ${P_s} = \frac{{2(\kappa  - 1)}}{\kappa }Q(\sqrt {\frac{{6{{\log }_2}(\kappa )}}{{{\kappa ^2} - 1}}(\frac{{{E_b}}}{{{N_0}}})} )$, where $\kappa  = \sqrt M$, $E_b$ is the 1 bit energy, $N_0$ is unilateral power spectral density of AWGN, and  $Q( \bullet )$ is the tail probability of the standard normal distribution, i.e. $Q(x) = \frac{1}{{\sqrt {2\pi } }}\int\limits_x^\infty  {{e^{ - \frac{{{u^2}}}{2}}}} du$.

However, this SER equation is confined to AWGN channel. In fading channels with channel gain $K_0$, the received 1 bit energy is $K_{0}E_b$. Therefore, in the case of OFDM with $M$-QAM over fading channel, the SER of the $i$th subcarrier is ${P_{s,i}} = \frac{{2(\kappa  - 1)}}{\kappa }Q(\sqrt {\frac{{6{{\log }_2}(\kappa )}}{{{\kappa ^2} - 1}}{{\left| {{H_i}} \right|}^2}(\frac{{{E_b}}}{{{N_0}}})} )$, where $H_i$ is the frequency-domain response of the $i$th subcarrier.

In terms of OFDM data block with DFT length $N$ and uncoded $M$-QAM, SER can be expressed as
\begin{align}
\!{P_s} = \frac{1}{N}\sum\limits_{n = 0}^{N - 1} {\frac{{2(\kappa  - 1)}}{\kappa }Q(\sqrt {{{\left| {\left. {H(n{f_0})} \right|} \right.}^2}\frac{{6{{\log }_2}(\kappa )}}{{{\kappa ^2} - 1}}(\frac{{{E_b}}}{{{N_0}}})} )},  \label{equ:SER}
\end{align}
where $f_0=1/N$.

If $M$-QAM adopts Gray map, BER can be approximated as
\begin{align}
{P_e} \approx {P_s}/{\log _2}(\kappa ).\label{equ:BER}
\end{align}
%, (\ref{equ:SER}), (\ref{equ:BER})
From (\ref{equ:H1})-(\ref{equ:BER}), it is clear that ${\left| {\left. {H(f )} \right|} \right.^2}$ has a great influence on SER or BER, and $H(f )$ can also be written as $H(f ;\varepsilon )$. Therefore, the optimal sampling phase $\varepsilon$ is to meet the minimum BER, i.e.
\begin{align}
{\varepsilon _{\text{opt}}} =\text{arg} \mathop {\min }\limits_\varepsilon  ({P_e}).\label{equ:e_optium}
\end{align}

Obviously, to obtain ${\varepsilon _{\text{opt}}}$ from (\ref{equ:e_optium}) is very difficult. Hence, we use Chernoff Bound \cite{BER_MQAM}, i.e.
\begin{align}
{P_e} \le \frac{1}{N}\sum\limits_{n = 0}^{N - 1} {\lambda \exp (-{{\left| {\left. {H(n{f_0};\varepsilon)} \right|} \right.}^2}\eta )},\label{equ:chenoff}
\end{align}
where $\lambda  = \frac{{2(\kappa  - 1)}}{\kappa }$ and $\eta  = \frac{{6{{\log }_2}(\kappa )}}{{{\kappa ^2} - 1}}\frac{{{E_b}}}{{{N_0}}}$.

We consider that in AWGN channel, according to (\ref{equ:HSRRC}) and (\ref{equ:H1}), $H(\Omega )$ can be written as
\begin{align}
\!\!\!\!&H(f)  \approx H_{\text{SRRC}}(f)=   \nonumber \\
 &\left\{ \begin{array}{l}
 {e^{j2\pi \varepsilon f }},{\kern 1pt} {\kern 1pt} {\kern 1pt} {\kern 1pt} {\kern 1pt} {\kern 1pt} {\kern 1pt} {\kern 1pt} {\kern 1pt} {\kern 1pt} {\kern 15pt} {\kern 1pt} {\kern 1pt} {\kern 1pt} {\kern 1pt} {\kern 1pt} {\kern 1pt} 0 \le f  < 0.5(1 - \alpha ), \\
 {e^{j2\pi \varepsilon (f  - 0.5)}}[\cos (\pi \varepsilon ) + \sin (\frac{\pi }{\alpha }(0.5 - f ))\sin (\pi \varepsilon )j],{\kern 1pt} {\kern 1pt} {\kern 1pt}  \\
{\kern 60pt} 0.5(1 - \alpha ) \le f  < 0.5(1 + \alpha ), \\
 {e^{j2\pi \varepsilon (f - 1)}},{\kern 1pt} {\kern 1pt} {\kern 1pt} {\kern 1pt} {\kern 1pt} {\kern 1pt} {\kern 1pt} {\kern 1pt} {\kern 1pt} {\kern 1pt} {\kern 1pt} {\kern 1pt}  0.5(1 + \alpha ) \le f  < 1, \\
 \end{array} \right.\label{equ:H2}
\end{align}

From (\ref{equ:H2}), it is obvious that the sampling phase offset $\varepsilon$ affects BER by affecting the ${\left| {\left. {H(\Omega ;\varepsilon)} \right|} \right.^2}$ of $\Omega  \in [0.5(1 - \alpha ),0.5(1 + \alpha )]$.
Thus a sampling phase criterion in AWGN channel can be acquired based on (\ref{equ:chenoff})
\begin{align}
\begin{array}{l}
\!\!\!\!\!\!{\varepsilon _{{\rm{AWGN}}}} = \text{arg} \mathop {\min }\limits_\varepsilon  (\sum\limits_{n = \left\lceil {0.5N(1 - \alpha )} \right\rceil }^{\left\lfloor {0.5N(1 + \alpha )} \right\rfloor } {\exp (-{{\left| {\left. {H(n{f _0})} \right|} \right.}^2}\eta )} ) \\
   {\kern 34pt} {\kern 1pt} {\kern 1pt} {\kern 1pt} {\kern 1pt} {\kern 1pt}  = \text{arg} \mathop {\min }\limits_\varepsilon  (\sum\limits_{n = \left\lceil {0.5N(1 - \alpha )} \right\rceil }^{\left\lfloor {0.5N(1 + \alpha )} \right\rfloor } {\exp [-(\cos {{(\pi \varepsilon )}^2} }  \\
 {\kern 1pt} {\kern 1pt} {\kern 1pt} {\kern 1pt} {\kern 1pt} {\kern 1pt} {\kern 1pt} {\kern 1pt} {\kern 1pt} {\kern 1pt} {\kern 1pt} {\kern 1pt} {\kern 1pt} {\kern 1pt} {\kern 1pt} {\kern 1pt} {\kern 1pt} {\kern 1pt} {\kern 1pt} {\kern 1pt} {\kern 1pt} {\kern 1pt} {\kern 1pt} {\kern 1pt} {\kern 1pt} {\kern 1pt} {\kern 1pt} {\kern 1pt} {\kern 1pt} {\kern 1pt} {\kern 1pt} {\kern 1pt} {\kern 1pt} {\kern 1pt} {\kern 1pt} {\kern 1pt} {\kern 1pt} {\kern 1pt} {\kern 1pt} {\kern 1pt} {\kern 1pt} {\kern 1pt} {\kern 1pt} {\kern 1pt} {\kern 1pt} {\kern 1pt} {\kern 1pt} {\kern 1pt} {\kern 1pt} {\kern 1pt} {\kern 1pt} {\kern 1pt} {\kern 1pt} {\kern 1pt} {\kern 1pt} {\kern 1pt} {\kern 1pt} {\kern 1pt} {\kern 1pt} {\kern 1pt} {\kern 1pt} {\kern 1pt} {\kern 1pt} + \sin {(\pi \varepsilon )^2}\sin {(\frac{\pi }{\alpha }(0.5 - n{f_0}))^2})\eta ]), \\
 \end{array}\label{equ:opt_awgn1}
 \end{align}
where $\left\lceil  \bullet  \right\rceil$ and $\left\lfloor  \bullet  \right\rfloor$ are integer ceiling and floor operators, respectively.

Furthermore, from $\frac{{\partial (\cos {{(\pi \varepsilon )}^2} + \sin {{(\pi \varepsilon )}^2}\sin {{(\frac{\pi }{\alpha }(0.5 - f))}^2})}}{{\partial \varepsilon }} = \pi \sin (2\pi \varepsilon )[\sin {(\frac{\pi }{\alpha }(0.5 - f))^2} - 1]$, we observe that $\cos {(\pi \varepsilon )^2} + \sin {(\pi \varepsilon )^2}\sin {(\frac{\pi }{\alpha }(0.5 - f ))^2}$ obtains the maximum value with $ \varepsilon={\varepsilon _{{\rm{AWGN}}}}=0$, which is because the value of the first-order partial derivative $\pi \sin (2\pi \varepsilon )[\sin {(\frac{\pi }{\alpha }(0.5 - f))^2} - 1]$ is positive when $ \varepsilon<0$, and negative when $ \varepsilon>0$. Therefore, (\ref{equ:opt_awgn1}) is optimal over AWGN channel and it can be also expressed by \begin{align}
\begin{array}{l}
 {\varepsilon _{{\rm{AWGN}}}} =\text{arg} \mathop {\max }\limits_\varepsilon  (\sum\limits_{n = \left\lceil {0.5N(1 - \alpha )} \right\rceil }^{\left\lfloor {0.5N(1 + \alpha )} \right\rfloor } {\cos {{(\pi \varepsilon )}^2} + }  \\
 {\kern 1pt} {\kern 1pt} {\kern 1pt} {\kern 1pt} {\kern 1pt} {\kern 1pt} {\kern 1pt} {\kern 1pt} {\kern 1pt} {\kern 1pt} {\kern 1pt} {\kern 1pt} {\kern 1pt} {\kern 1pt} {\kern 1pt} {\kern 1pt} {\kern 1pt} {\kern 1pt} {\kern 1pt} {\kern 1pt} {\kern 1pt} {\kern 20pt} {\kern 1pt} {\kern 1pt} {\kern 1pt} {\kern 1pt} {\kern 1pt} {\kern 1pt} {\kern 1pt} {\kern 1pt} {\kern 1pt} {\kern 1pt} {\kern 1pt} {\kern 1pt} {\kern 1pt} {\kern 1pt} {\kern 1pt} {\kern 1pt} {\kern 1pt} {\kern 1pt} {\kern 1pt} {\kern 1pt} \sin {(\pi \varepsilon )^2}\sin (\frac{\pi }{\alpha }{(0.5 - n{f_0})^2}). \\
 \end{array}\label{equ:opt_awgn2}
\end{align}

Compared with (\ref{equ:e_optium}) and (\ref{equ:opt_awgn1}), (\ref{equ:opt_awgn2}) is more feasible, which only needs the sum of partial squared channel frequency-domain gains. Moreover, this sampling phase criterion can also be extended to all channel conditions, i.e.
\begin{align}
{\varepsilon _{{\rm{general}}}} =\text{arg} \mathop {\max }\limits_\varepsilon  (\sum\limits_{n = \left\lceil {0.5N(1 - \alpha )} \right\rceil }^{\left\lfloor {0.5N(1 + \alpha )} \right\rfloor } {{{\left| {H(n{f_0})} \right|}^2}} ).\label{equ:opt_awgn3}
\end{align}
(\ref{equ:opt_awgn3}) may not be optimal over multipath channels, but it is near optimal and its validity will be demonstrated in Section IV.

In a sense, the correction of the sampling phase offset is a more fine synchronization operation. Therefore, this correction is implemented after the receiver achieves the perfect frame synchronization, CFO correction and sampling frequency offset elimination. Consequently, in practical application, the sampling phase correction module should be cascaded following the synchronization module shown in Fig. \ref{fig:baseband}.
\begin{figure}[tblp]
     \centering
     \includegraphics[width=8.0cm, keepaspectratio]
     {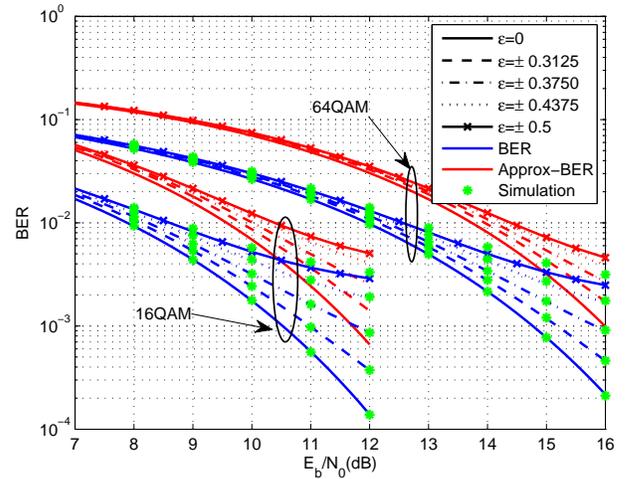}
     \vspace*{-2mm}
     \caption{BER of theoretical analysis, approximated BER, and BER of simulation with different sampling phase offsets over AWGN channel. Uncoded 16QAM and 64QAM are used for OFDM data symbols.}
     \label{fig:the_vs_sim}
     \vspace*{-5mm}
\end{figure}
\section{Simulation Results}
This section investigates the performances of the proposed sampling phase offset estimation criterion and the conventional STR method \cite{FS0}. Simulations assume perfect frame synchronization, sampling clock frequency offset recovery, and CFO elimination. Dual PN-OFDM (DPN-OFDM) is adopted for TDS-OFDM transmission with OFDM data block length $N=4096$, and PN sequence length $L=512$. Upsampling factor and SRRC roll-off factor are the same with DTMB system, i.e. $N_{\text{upsam}}=4$, $\alpha=0.05$. Additionally, the Brazil digital television field test 4th (Brazil-B) and 5th (Brazil-E) channel models \cite{channel} are selected.

Fig. \ref{fig:the_vs_sim} shows the BERs of theoretical analysis based on (\ref{equ:SER}), (\ref{equ:BER}), (\ref{equ:H2}) and the BERs of simulation with different sampling phase offsets ($\varepsilon = 0,\pm 0.3125, \pm 0.3750, \pm 0.4375, \pm 0.5$) over AWGN channel. Uncoded 16QAM and 64QAM are used for OFDM data symbols. Additionally, the approximated BERs using Chernoff Bound are also plotted for comparison, denoted as Approx-BER. This figure strongly verifies the validity of the BER of theoretical analysis in Section \ref{proposed}. The performance curves obtained via the theoretical approach are in good agreement with that of the Monte Carlo simulation results. Meanwhile, we also observe that Approx-BER is a good approximation of BER, since curves of Approx-BER are the horizontal axis shift versions of curves of BER. Therefore, these powerfully support the rationality of the sampling phase estimation criterion derived from the approximated BER using Chernoff Bound.

In Fig. \ref{fig:the_vs_sim}, it is obvious that, with the sampling phase offset increasing, the BER performance degrades rapidly. Here, the target BER of $3 \times {10^{{\rm{ - }}3}}$ is considered. To achieve the target BER, the best BER performance is superior to the worst situation by 2.5dB performance gain. Moreover, with $E_b/N_0$ increasing, the BER performance differences of different sampling phases increase. Furthermore, from the BER curve tendency of different sampling phases, it can be observed that with the sampling phase increases, the BER floor phenomenon is more obvious.
\begin{figure}[tblp]
     \centering
     \includegraphics[width=8.0cm, keepaspectratio]
     {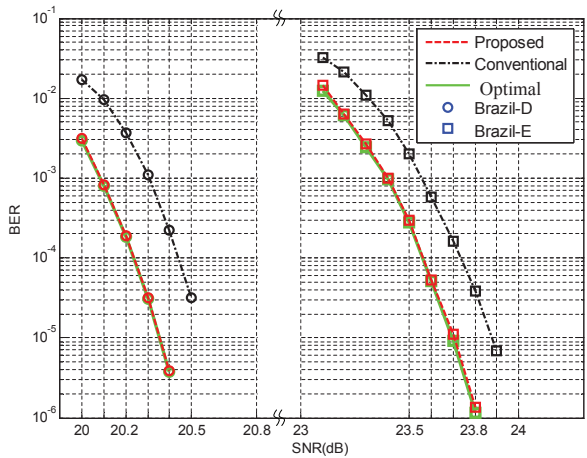}
     \vspace*{-2mm}
     \caption{BER performance comparison when 256QAM is adopted over multipath channels.}
     \label{fig:256qam}
     \vspace*{-5mm}
\end{figure}

Fig. \ref{fig:256qam} compares the low-density parity check (LDPC) coded BER performance of the proposed criterion and the conventional STR method, where 256QAM is adopted and LDPC code rate is 0.6. Additionally, we adopts the grid search method to approach the BER of the optimal sampling phase. In this method, we divide the sampling period $T_{\text{sym}}$ into 128 uniformly-spaced sampling phases, and consider the best performance of BERs associated with different sampling phases as the BER of the optimal sampling phase. In Fig. \ref{fig:256qam}, the proposed sampling phase estimation criterion has superior BER performance to its counterpart. The performance gain can be 0.2dB over both Brazil-D and Brazil-E channels. In addition, the proposed criterion performs closely to the optimal sampling phase, which indicates the excellent performance of the proposed criterion. Therefore, the proposed method can be considered to be near optimal.
%\begin{figure}[tblp]
%     \centering
%     \includegraphics[width=9.5cm, keepaspectratio]
%     {figs/ldpc_ber64.eps}
%     \caption{Theoretical analysis BER, approximate BER, and simulation BER of different sampling phase offsets in uncoded 16QAM modulation or 64QAM over AWGN channel.}
%     \label{fig:the_vs_sim}
%\end{figure}

\section{Conclusion}
In this paper, we derived the theoretical BER for uncoded TDS-OFDM system over AWGN channel, and it can be observed that the sampling phase offset has a great impact on BER. Furthermore, we proposed a near optimal sampling phase estimation criterion. Compared with the conventional time-domain based synchronization methods for TDS-OFDM, the proposed criterion obtains the near optimal sampling phase based on the frequency-domain response. Simulations demonstrate that the proposed criterion has good performance in actual LDPC-coded TDS-OFDM system over multipath channels. The proposed criterion is superior to the conventional STR method and performs closely to the optimal sampling phase over multipath channels.
%\section{Acknowledgments}
%This work was supported by Program for New Century
%Excellent Talents in University and National High Technology Research and Development Program of China (Grant No. 2012AA011704), %National Nature Science Foundation of China (Grant No. 60902003).
\section{Acknowledgment}
This work was supported in part by the National Nature Science Foundation of China (Grant No. 61271266) and the open project ITD-U1300x/K13600xx of Science and Technology on Information Transmission and Dissemination in Communication Networks Laboratory.
%\vspace{-0.2cm}

\end{document}